\documentclass{article} 
\usepackage{amsmath,amssymb,amsbsy,amsfonts}
\makeatletter

          \@addtoreset{equation}{section}
       \makeatother
\allowdisplaybreaks 
\begin{document}
\title{\bf Lorentz invariant nonabelian gauge theory 
on noncommutative space-time 
and BRST symmetry }%

\author{Yoshitaka {\sc Okumura}$^\dag$, 
Katsusada {\sc Morita}$^\ddag$, Kouhei {\sc Imai}$^\dag$\\
{$^\dag$Department of Natural Science, 
Chubu University, Kasugai, {487-8501}, Japan}\\
$^\ddag${Department of Physics, Nagoya University, 
Nagoya, {464-8602}, Japan}
}%
\date{}%
\maketitle
\begin{abstract}
Lorentz covariance is the fundamental principle of every relativistic field theory which insures consistent physical descriptions. Even if the space-time is noncommutative, field theories on it should keep Lorentz covariance.
In this paper, the nonabelian gauge theory on noncommutative spacetime is defined and its Lorentz invariance is maintained based on the idea of 
Carlson, Carone and Zobin.
The deviation from the standard model in particle physics has not yet observed,
and so any model beyond standard model must reduce to it in 
some approximation. Noncommutative gauge theory must also reproduce standard
model in the limit of noncommutative parameter $\theta^{\mu\nu}\to0$. 
Referring to Jur$\check{\text{c}}$o {\it et. al.}, we 
will construct the nonabelian gauge theory 
that deserves to formulate standard model.
BRST symmetry is very important to quantize nonabelian gauge theory
and construct the covariant canonical formulation. 
It is discussed about the fields in noncommutative gauge theory 
without considering those components.
Scale symmetry of ghost fields is also discussed.
\end{abstract}



\thispagestyle{empty}
\section{Introduction\label{S1}}
In the past several years, field theories on the noncommutative spacetime
have been extensively studied from many different aspects. The motivation 
comes from the string theory which makes obvious that end points of 
the open strings trapped on the D-brane in the presence of 
 two form B-field background turn out to be noncommutative 
\cite{SJ}
and 
then the noncommutative supersymmetric gauge theories appear as the low energy effective theory of such D-brane \cite{DH}, \cite{SW}.
Though this is a driving force of recent prevalence of 
noncommutative field theories, noncommutative spacetime 
has long history. Especially, Snyder \cite{SN} proposed 
Lorentz invariant algebra between spacetime ${\hat x}^\mu$ and 
the generator of Lorentz transformation $\hat{M}^{\mu\nu}$ and 
showed the existence of a noncommutative spacetime with a fundamental length.
Based on the Snyder's algebra, the relativistic field theory was developed
 \cite{Kad} though it was not so successful.
The noncommutativity of spacetime in recent surge 
is characterized by the algebra
$[x^\mu,\,x^\nu]=i\theta^{\mu\nu}$ where $\theta^{\mu\nu}$ 
is a real, anti-symmetric constant with dimension 2, which is reflected as the Moyal star product in field theories. 
According to this prescription, one can build noncommutative version of 
scalar, Dirac and gauge theories. Thus, apart from the string theory,
the studies of noncommutative field theories have been proposing
very interesting as well as reversely serious outcomes.
The noncommutative scalar field theories are investigated in \cite{Filk}, 
\cite{SST},\cite{Micu}, \cite{GM}.
It was showed that the noncommutative scalar theory with $\mit\Phi^4$
interaction is renormalizable and the parameter $\theta^{\mu\nu}$ 
doesn't receive the quantum corrections up to two loop order. 
However, since the Moyal star product contains an infinite series of 
$x^\mu$ derivatives, noncommutative field theories are nonlocal.
The nonlocality especially in timelike noncommutativity $\theta^{0i}\ne0$ 
leads to the unitarity violation \cite{GM} and the difficulty of 
renormalizability. 
All these investigations are 
carried out under the condition of constant $\theta^{\mu\nu}$, 
which means that Lorentz covariance of theory is violated. \par
Doplicher, Fredenhagen and Roberts (DFR) proposed \cite{DFR} a new algebra of 
noncommutative spacetime through consideration of the spacetime uncertainty
relations derived from quantum mechanics and general relativity.
In their algebra, $\theta^{\mu\nu}$ is promoted to an anti-symmetric tensor operator, which leads to the Lorentz covariant noncommutative spacetime and
enables one to construct the Lorentz invariant noncommutative field theories.
Carlson, Carone and Zobin (CCZ) \cite{CCZ} 
formulated the noncommutative gauge theory by referring to the DFR algebra
in the Lorentz invariant way. In their formulation, fields in the theory
depend on spacetime $x^\mu$ and the noncommutative parameter
$\theta^{\mu\nu}$. The action is obtained by integrating Lagrangian 
over spacetime $x^\mu$  as well as the noncommutative parameter
$\theta^{\mu\nu}$.
The characteristic features of CCZ \cite{CCZ} are to set up the 6-dimensional
$\theta$ space in addition to $x$ space and define the action as the integration of Lagrangian over the $\theta$ space as well as $x$ space.
In this article, we will construct Lorentz invariant noncommutative field theory by  taking over the idea of the 6-dimensional $\theta$ space, but
not  the integration over $\theta$ space. 
We can choose the spacelike noncommutativity ($\theta^{0i}=0$) after 
the appropriate Lorentz transformation and so any serious outcomes such as 
quantization and unitarity violation don't appear.
\par
U(1) noncommutative gauge theory requires the matter field to have charge 
0 or $\pm1$ in order to keep the gauge invariance and define the covariant derivative \cite{Hayakawa}. Armoni \cite{Armoni} indicated that U(N) gauge theory has the consistency in calculations of gluon propagator and  three gluons vertex to one loop order, whereas SU(N) gauge theory is not consistent. These problems have been overcome by Jur$\check{\text{c}}$o, {\it et al} Jurco who constructed nonabelian
gauge theory on the enveloping Lie algebra steming from the Moyal star product. No extra fields other than fields in ordinary commutative gauge field theory 
appear in their formulation after performing the Seiberg-Witten map.
This approach has allowed them to construct the noncommutative standard model
as well as SO(10) GUT \cite{Jurco2}. Referring to their consideration, we will
formulate the noncommutative nonabelian gauge theory by use of the enveloping
group SU(N)$\ast$ with enveloping Lie algebra. \par
BRST symmetry is very important to quantize nonabelian gauge theory
and construct the covariant canonical formulation of it \cite{KO}.
This subject in noncommutative  U(N) gauge theory was recently 
studied by Soroush \cite{SOR} by determining the BRST transformations
of the components of gauge and related fields. Let us in this article 
introduce the BRST transformations of fields 
as the representations of gauge group, 
not component of fields, which
leads us to transparent view of BRST symmetry.
The scale transformation of ghost fields under which full Lagrangian of gauge theory is invariant is also discussed. The algebra of BRST charge and FP ghost charge which may serve the classification of asymptotic fields and the proof of S matrix unitarity is displayed.
\par

\section{Nonabelian gauge theory on noncommutative spacetime\label{S2}}
Let us first consider the nonabelian gauge theory on the commutative spacetime
with the symmetry SU(N)  given by the Lagrangian
\begin{align}
{\cal L}=-\frac14\text{Tr}\left[F_{\mu\nu}(x)F^{\mu\nu}(x)\right]
+{\mit\bar\Psi}(x)\{i\gamma^\mu(\partial_\mu-igA_\mu(x))-m\}{\mit\Psi}(x),
\label{2.1}
\end{align}
where we omit the gauge fixing and FP ghost terms and
\begin{equation}
F_{\mu\nu}(x)=\partial_\mu A_\nu(x)-\partial_\nu A_\mu(x)-ig\,[A_\mu(x),\,A_\nu(x)]\label{2.2}
\end{equation}
with the configuration
\begin{equation}
A_\mu(x)=\sum_{a=1}^{N^2-1}A^a_\mu(x)T^a.\label{2.3}
\end{equation}
Lagrangian \eqref{2.1} is invariant under the gauge transformations
\begin{align}
& A^g_\mu(x)=U(x)A_\mu(x)U^{-1}(x)+\frac{i}{g}U(x)\partial_\mu U^{-1}(x),\label{2.41}\\
&{\mit\Psi}^g(x)=U(x){\mit\Psi}(x),\label{2.42}
\end{align}
where the gauge transformation function $U(x)$ is written as
\begin{equation}
U(x)=e^{i\alpha(x)}
\end{equation}
with the Lie algebra valued function
\begin{equation}
\alpha(x)=\sum_{a=1}^{N^2-1}\alpha^a(x)T^a.
\end{equation}
The commutator between two Lie algebra valued functions 
$\alpha(x)$ and $\beta(x)$ 
\begin{align}
[\alpha(x),\,\beta(x)]=&\sum_{a,\,b=1}^{N^2-1}\alpha^a(x)\beta^b(x)\left[T^a,\,
T^b\right]\nonumber\\
=&\sum_{c=1}^{N^2-1}\left(\sum_{a,b=1}^{N^2-1}if^{abc}\,\alpha^a(x)\beta^b(x)\right)T^c
\label{2.8}
\end{align}
is also Lie algebra valued function. 
Thus, they constitute the closed Lie algebra.

\par
Then, we introduce the nonabelian gauge theory 
on the noncommutative spacetime by
employing the Moyal $\ast$product
\begin{equation}
f(x)\ast g(x)=\left.e^{\frac{i}{2}\theta^{\mu\nu}\partial^1_\mu\partial^2_\nu}f(x_1)g(x_2)\right|_{x_1=x_2=x},
\end{equation}
where $\theta^{\mu\nu}$ is two rank tensor 
to characterize the noncommutativity of
spacetime.  Though  $\theta^{\mu\nu}$ is usually seemed to be 
a constant not to transform corresponding to Lorentz transformation,
$\theta^{\mu\nu}$ is regarded as  two rank 
tensor in this article as discussed in the next section.
Because of the Moyal $\ast$product, $\ast$commutators between 
the Lie algebra valued functions don't 
close within themselves and extend to the enveloping Lie algebra since
Eq. \eqref{2.8} changes to 
\begin{align}
[\alpha(x),\,\beta(x)]_{\ast}=&\sum_{a,\,b=1}^{N^2-1}
\left(\alpha^a(x)\ast\beta^b(x)T^aT^b-\beta^b(x)\ast\alpha^a(x)T^b
T^a\right)
\nonumber\\
=&\sum_{c=1}^{N^2-1}\sum_{a,b=1}^{N^2-1}\left(if^{abc}\,\frac12\{\alpha^a(x),\,\beta^b(x)\}_{\ast}T^c+d^{abc}\,\frac12[\alpha^a(x),\,\beta^b(x)]_{\ast}\right)T^c
\nonumber\\
&+\sum_{a,b=1}^{N^2-1}d^{ab0}\,\frac12[\alpha^a(x),\,\beta^b(x)]_{\ast}T^0,
\label{2.10}
\end{align}
where
\begin{equation}
[T^a,\,T^b]=\sum_{c=1}^{N^2-1}if^{abc}T^c,\hskip1cm
\{T^a,\,T^b\}=\sum_{c=0}^{N^2-1}d^{abc}T^c.
\end{equation}
Thus, in general, 
the enveloping Lie algebra valued functions are written as
\begin{equation}
\alpha(x,\theta)=\sum_{a=0}^{N^2-1}\alpha^a(x,\theta)T^a,\label{2.12}
\end{equation}
where the condition
\begin{equation}
\lim_{\theta\to0}\alpha^0(x,\theta)=0
\label{2.13b}
\end{equation}
should be satisfied according to Eq.\eqref{2.10}.
In terms of the enveloping Lie algebra, we are able to construct the 
enveloping nonabelian group SU(N)$\ast$ which elements are defined by
\begin{equation}
U(x,\theta)=e^{i\alpha(x,\theta)}
\end{equation}
with Eq.\eqref{2.12}. It should be noted that 
\begin{equation}
\lim_{\theta\to0}U(x,\theta)=U(x)=e^{i\sum_{a=1}^{N^2-1}\alpha^a(x)T^a}
\in \text{SU(N)},\label{2.15}
\end{equation}
which indicates that the enveloping group SU(N)$\ast$ reduces to nonabelian
group SU(N) when $\theta^{\mu\nu}$ becomes to 0.
\par
Lagrangian of the nonabelian gauge theory 
on the noncommutative spacetime 
 is simply 
obtained by replacing the ordinary product with the Moyal $\ast$product
in Eq.\eqref{2.1},  
\begin{align}
{\cal L}=-\frac14\text{Tr}\left[F_{\mu\nu}(x)\ast F^{\mu\nu}(x)\right]
+{\mit\bar\Psi}(x)\ast\{i\gamma^\mu {\cal D}_\mu-m\}\ast{\mit\Psi}(x).
\label{2.16}
\end{align}
where ${\cal D}_\mu=\partial_\mu-igA_\mu(x)$ is covariant derivative and
\begin{equation}
F_{\mu\nu}(x)=\partial_\mu A_\nu(x)-\partial_\nu A_\mu(x)
-ig\,[A_\mu(x),\,A_\nu(x)]_\ast
\label{2.17a}
\end{equation}
with the definition of $\ast$commutator 
\begin{equation}
[A(x),\,B(x)]_\ast=A(x)\ast B(x)-B(x)\ast A(x).
\end{equation}
Gauge transformations of the fields are given by
\begin{align}
& A^g_\mu(x)=U(x,\theta)\ast A_\mu(x)\ast U^{-1}(x,\theta)+\frac{i}{g}U(x,\theta)\ast \partial_\mu U^{-1}(x,\theta),\label{2.17}\\
&{\mit\Psi}^g(x)=U(x,\theta)\ast {\mit\Psi}(x),\label{2.18}
\end{align}
Owing to the algebraic rules of Moyal $\ast$product, $F_{\mu\nu}(x)$ and 
${\cal D}_\mu{\mit\Psi(x)}$ 
are transformed covariantly
\begin{align}
&F_{\mu\nu}^g(x)=U(x,\theta)\ast F_{\mu\nu}(x)\ast U^{-1}(x,\theta),\label{2.19}\\
&\left\{{\cal D}_\mu{\mit\Psi(x)}\right\}^g=U(x,\theta)\ast {\cal D}_\mu{\mit\Psi(x)}\label{2.20}
\end{align}
Under the gauge transformation of $F_{\mu\nu}$ in Eq.\eqref{2.19}, 
the gauge field term in Eq.\eqref{2.16} is transformed as in
\begin{equation}
\text{Tr}\left[F_{\mu\nu}^g(x)\ast {F^{\mu\nu}}^g(x)\right]
=\text{Tr}\left[U(x,\theta)\ast F_{\mu\nu}(x)\ast F^{\mu\nu}(x)\ast U^{-1}(x,\theta)\right]
\label{2.22a}
\end{equation}
which shows the gauge term itself is not gauge invariant because of the 
Moyal $\ast$product but the action is invariant thanks to the rule
\begin{equation}
\int d^4x \;f(x)\ast g(x)=\int d^4x \;g(x)\ast f(x).
\end{equation}
On the other hand, the fermion term in Eq.\eqref{2.16} is invariant under gauge transformations  Eqs.\eqref{2.18} and \eqref{2.20}.
\par
In order to construct the nonabelian gauge theory on the noncommutative 
spacetime we start from Eq.\eqref{2.16} where the gauge field $A_\mu(x)$
doesn't contain the 0 component $A_\mu^0(x)$, but
it is induced in the gauge transformation.
However, the 0 component of ${A^g}_\mu(x)$ in Eq.\eqref{2.17}
depends on  $A_\mu(x)$ in Eq.\eqref{2.16}
and so, it is not independent field. 
Moreover it vanishes owing to Eq.\eqref{2.15} 
when the noncommutative parameter $\theta^{\mu\nu}$
approaches to 0. 
Thus, even if the 0 component of  ${A^g}_\mu(x)$ appears
in Eq.\eqref{2.17}, it is out of our considerations.
The existing of the enveloping SU(N)$\ast$ group
insures the construction of nonabelian gauge theory on the noncommutative 
spacetime.
\section{Lorentz invariance of noncommutative field theory}
Lorentz invariance is the fundamental principle of every relativistic field theory which insures consistent physical descriptions such as causality, 
unitarity and so on. However, it hasn't been respected in the study of 
noncommutative field theory so far. Doplicher, Fredenhagen and Roberts (DFR)
\cite{DFR}
first addressed this problem to propose a new algebra of noncommutative
spacetime operator ${\hat x}_\mu$.
\begin{equation}
[{\hat x}^\mu,\,{\hat x}^\nu]=i\,{\hat \theta}^{\mu\nu},
\end{equation}
where ${\hat \theta}^{\mu\nu}$ is an antisymmetric tensor operator, not a constant considered so far. They further assumed 
\begin{equation}
[{\hat x}^\mu,\,{\hat \theta}^{\mu\nu}]=0,
\end{equation}
which leads to the commutativity between ${\hat \theta}^{\mu\nu}$
through the Jacobi identity
\begin{equation}
[{\hat \theta}^{\mu\nu},\,{\hat \theta}^{\sigma\rho}]=0.
\label{3.3}
\end{equation}
Equation \eqref{3.3} enables us to simultaneously diagonize the operator
${\hat \theta}^{\mu\nu}$.
\begin{equation}
{\hat \theta}^{\mu\nu}|\,\theta>={\theta}^{\mu\nu}|\,\theta>,
\label{3.4}
\end{equation}
where $|\,\theta>$ is a eigenstate and ${\theta}^{\mu\nu}$ is 
its specific eigenvalue.\par
Carlson, Carone and Zobin (CCZ) \cite{CCZ} 
formulated the noncommutative gauge theory by referring to the DFR algebra
in the Lorentz invariant way. In their formulation, fields in the theory
depend on spacetime $x^\mu$ and the noncommutative parameter
$\theta^{\mu\nu}$. The action is obtained by integrating Lagrangian 
over spacetime $x^\mu$  as well as the noncommutative parameter
$\theta^{\mu\nu}$.
\begin{equation}
S=\int d^{\,4}x \,d^{\,6}\theta\; 
W(\theta){\cal L}(\phi(x,\theta),\partial^\mu 
\phi(x,\theta)),
\end{equation}
where Lorentz invariant function $W(\theta)$ is a weight function to render the $\theta$ integral finite. Following to CCZ, Kase, Morita, Okumura and Umezawa 
\cite{KMOU} reconsidered the Lorentz invariant noncommutative field theory by  
pointing out the inconsistency of the c-number $\theta$-algebra
 and indicated that
the normalizability of the weight
function in Lorentz metric leads to the division of the $\theta$ space 
into two disjoint regions not connected by any Lorentz transformation
, so that the CCZ covariant moments formula holds in each region separately.
\par
The characteristic features of CCZ \cite{CCZ} are to set up the 6-dimensional
$\theta$ space in addition to $x$ space and define the action as the integration of Lagrangian over the $\theta$ space as well as $x$ space.
In this article, we take over the idea of the 6-dimensional $\theta$ space, but
don't  the integration over $\theta$ space. Let us pick up one specific point 
$\theta^{\mu\nu}$ in 
the 6-dimensional $\theta$ space that follows from Eq.\eqref{3.4}.
If we denote the Lorentz transformation operator to be 
$U(\Lambda)$, the equation
\begin{equation}
U(\Lambda){\cal L}({\hat x},\;{\hat\theta}^{\mu\nu})U^{-1}(\Lambda)=
{\cal L}({\hat x}{\,'},\;{\hat\theta}^{\,'}\rule{0mm}{2.8mm}^{\mu\nu})
\end{equation}
holds. Since the Lagrangian is invariant under Lorentz transformation, 
the nontrivial equation
\begin{equation}
{\cal L}({\hat x},\;{\hat\theta}^{\mu\nu})=
{\cal L}({\hat x}{\,'},\;{\hat\theta}^{\,'}\rule{0mm}{2.8mm}^{\mu\nu})
\label{3.7}
\end{equation}
follows. 
In order to obtain the Lorentz invariant Lagrangian from this equation, 
we derive several usuful equations.
When the Lorentz transformation operator $U(\Lambda)$ works on Eq.\eqref{3.4}
the equation
\begin{equation}
U(\Lambda)\,{\hat\theta}^{\,\mu\nu}U^{-1}(\Lambda)\,U(\Lambda)\,|\,\theta>
=\,{\hat\theta}^{\,'}\rule{0mm}{3mm}^{\mu\nu}\,|\,\theta^{'}>
={\theta}^{\mu\nu}\,|\,\theta^{'}>
\end{equation}
 follows, where
\begin{equation}
\begin{aligned}
|\,\theta^{'}>&=U(\Lambda)\,|\,\theta>,\hskip1cm
<\theta^{'}\,|=<\theta\,|\,U^{-1}(\Lambda),\\
{\hat\theta}^{\,'}\rule{0mm}{3mm}^{\mu\nu}&=U(\Lambda)\,{\hat\theta}^{\,\mu\nu}\,U^{-1}(\Lambda).
\end{aligned}
\end{equation}
Since the Lorentz transformation for operator ${\hat\theta}^{\,\mu\nu}$ is
\begin{equation}
U(\Lambda) {\hat\theta}^{\,\mu\nu}U^{-1}(\Lambda)=
\Lambda^{\;\,\mu}_{\rho}\Lambda^{\;\,\nu}_{\sigma}
{\hat\theta}^{\,\rho\sigma}=\,{\hat\theta}^{\,'}\rule{0mm}{3mm}^{\mu\nu}
\label{3.10}
\end{equation}
the Lorentz transformation for its eigenvalue ${\theta}^{\,\mu\nu}$ is
\begin{equation}
\Lambda^{\;\,\mu}_{\rho}\Lambda^{\;\,\nu}_{\sigma}
{\theta}^{\,'}\rule{0mm}{3mm}^{\,\rho\sigma}=\,{\theta}^{\mu\nu}
\hskip3mm\Longleftrightarrow\hskip3mm
\Lambda^{\mu}_{\;\,\rho}\Lambda^{\nu}_{\;\,\sigma}
{\theta}^{\,\rho\sigma}=\,{\theta}^{\,'}\rule{0mm}{3mm}^{\mu\nu}.
\end{equation}
where ${\theta}^{\,'}\rule{0mm}{3mm}^{\mu\nu}$ is defined as 
\begin{equation}
 {\hat\theta}^{\,\mu\nu}\,|\,\theta^{'}>
=\,{\theta}^{\,'}\rule{0mm}{3mm}^{\mu\nu}\,|\,\theta^{'}>.
\end{equation}
Operating $U^{-1}(\Lambda)$ on the above equation and using Eq.\eqref{3.10},
we derive
\begin{equation}
{\hat\theta}^{\,'}\rule{0mm}{3mm}^{\mu\nu}\,|\,\theta>
=\,{\theta}^{\,'}\rule{0mm}{3mm}^{\mu\nu}\,|\,\theta>.
\label{3.14}
\end{equation}
Sandwiching Eq.\eqref{3.7} with $|\,\theta>$, we obtain the equation 
\begin{align}
<\theta\,|\,{\cal L}({\hat x},\,{\hat\theta})^{\mu\nu}\,|\,\theta>
=<\theta\,|\,{\cal L}({\hat x}{\,'},\;{\hat\theta}^{\,'}\rule{0mm}{2.8mm}^{\mu\nu})\,|\,\theta>
\end{align}
which owing to Eq.\eqref{3.14} leads to 
\begin{align}
\,{\cal L}({\hat x},\,{\theta}\rule{0mm}{2.8mm}^{\mu\nu})\,
={\cal L}({\hat x}{\,'},\;{\theta}^{\,'}\rule{0mm}{2.8mm}^{\mu\nu}).
\label{3.6}
\end{align}
after the normalization factor
$<\theta\,|\,\theta>$ is scaled out.

\par
In this stage, the choice of $\theta^{\,\mu\nu}$ is arbitrary. 
However, quantization restricts the allowable region of $\theta^{\,\mu\nu}$
because when the timelike noncommutativity $\theta^{0i}\ne0$ exists,
the conjugate momentum of a filed $\phi$ defined by
\begin{equation}
{\Pi}=\frac{\partial {\cal L}}{\partial(\partial^0 \phi)}
\end{equation}
is not qualified as an appropriate momentum owing to the infinite
series of time derivatives in the Moyal $\ast$products.
Thus, we can restrict the region of $\theta^{\,\mu\nu}$ in such a way that
we can render 
$\theta^{\,0i}$ to be 0 by making an appropriate Lorentz transformation of 
$\theta^{\,\mu\nu}$.
The unitarity problem pointed out by Gomis and Mehen \cite{GM}
might vanish by considering the Lorentz invariance of the theory 
and the proper choice of $\theta^{\,\mu\nu}$ as discussed above. 
This problem will be elaborated in another article.
\section{BRST symmetry of noncommutative gauge theory}
BRST symmetry is very important to quantize the nonabelian gauge theory 
through which the physical states are normally defined. It also plays 
an indispensable role in deriving the Ward-Takahashi identity used to prove 
the renormalization of nonabelian gauge theory. Similarly, it could be 
important in the case of the nonabelian gauge theory 
on noncommutative spacetime. This subject was recently 
studied by Soroush \cite{SOR} by determining the BRST transformations
of the components of gauge and related fields. Let us here introduce the 
BRST transformations of fields as the representations of gauge group, 
not component of fields, which
leads us to transparent view of BRST symmetry.\par
The total Lagrangian of noncommutative gauge theory consists of
\begin{equation}
{\cal L}={\cal L}_G+{\cal L}_D+{\cal L}_{GF}+{\cal L}_{FP},
\label{4.1}
\end{equation}
where
\begin{align}
&{\cal L}_G=-\frac14\text{Tr}\left[F_{\mu\nu}(x)\ast F^{\mu\nu}(x)\right],\\
&{\cal L}_D={\bar{\mit\Psi}}(x)\ast \{i\gamma^\mu(\partial_\mu-ig A_\mu(x))-m\}
\ast{\mit\Psi}(x)\nonumber\\
&\hskip5mm={\bar{\mit\Psi}}(x)\ast \{i\gamma^\mu{\cal D}_\mu-m\}
\ast{\mit\Psi}(x),\\
&{\cal L}_{GF}=\text{Tr}\left(-\partial^\mu B(x)\ast A_\mu(x)+\frac{\alpha}{2}
B(x)\ast B(x)\right),\\
&{\cal L}_{F\!P}
=\text{Tr}\left\{-i\partial^\mu{\bar c}(x)\ast (\partial_\mu c(x)
-ig[A_\mu(x),\,c(x)]_\ast)\right\}\nonumber\\
&\hskip7mm=\text{Tr}\left\{-i\partial^\mu{\bar c}(x)\ast D_\mu c(x)\right\}.
\end{align}
${\cal L}_G$ is the gauge field term with $F_{\mu\nu}(x)$ given 
in Eq.\eqref{2.17}. 
${\cal L}_D$ is the Dirac fermion term in the 
fundamental representation of the gauge group. ${\cal L}_{GF}$
is the gauge fixing term with the auxiliary field $B(x)$ called 
the Nakanishi-Lautrup field. 
${\cal L}_{FP}$ is the Faddeev-Popov ghost term.
Ghost fields $c(x),\,{\bar c}(x)$ and Nakanishi-Lautrup field $B(x)$ are
denoted with components fields by 
\begin{align}
&c(x)=\sum_{a=1}^{N^2-1}c^a(x)T^a,\label{4.6}\\
&{\bar c}(x)=\sum_{a=1}^{N^2-1}{\bar c}^a(x)T^a,\label{4.7}\\
&B(x)=\sum_{a=1}^{N^2-1}B^a(x)T^a.\label{4.8a}
\end{align}
\par
BRST transformations of gauge field $A_\mu(x)$ and fermion field 
${\mit\Psi}(x)$ are derived by replacing $\alpha(x,\theta)$ 
by $ig\lambda c(x)$ in Eqs.\eqref{2.17} and \eqref{2.18}, respectively 
\begin{align}
&\delta_{\text{\tiny$B$}}A_\mu(x)=\lambda 
\left(\partial_\mu c(x)-ig[A_\mu(x),\,c(x)]_\ast\right)=\lambda D_\mu c(x),\\
&\delta_{\text{\tiny$B$}}{\mit\Psi}(x)=ig\lambda c(x)\ast {\mit\Psi}(x),
\end{align}
where the parameter $\lambda$ is a Grassmann variable
not depending on spacetime coordinate $x_\mu$. This parameter $\lambda$
 is usually eliminated in the formula. Such BRST transformation is written by
$\boldsymbol\delta_{\text{\tiny$ B$}}$ which satisfies the rule
\begin{equation}
\boldsymbol\delta_{\text{\tiny$ B$}}(F\ast G)=
(\boldsymbol\delta_{\text{\tiny$ B$}}F)\ast G+(-1)^{|F|}F\ast 
(\boldsymbol\delta_{\text{\tiny$ B$}}G),\label{4.8}
\end{equation}
where $|F|$ is the ghost number of field $F$.
\par
The BRST transformations of fields $c(x),\,{\bar c}(x)$, and $B(x)$ are
determined in order for the BRST operator $\boldsymbol\delta_{\text{\tiny$ B$}}$ to satisfy the nilpotency such as
\begin{align}
& \boldsymbol\delta_{\text{\tiny$ B$}}c(x)=igc(x)\ast c(x), \\
& \boldsymbol\delta_{\text{\tiny$ B$}}{\bar c}(x)=i B(x),\\
& \boldsymbol\delta_{\text{\tiny$ B$}}B(x)=0.
\end{align}
Let us show the nilpotency of 
the BRST operator $\boldsymbol\delta_{\text{\tiny$ B$}}$, that is 
$\boldsymbol\delta_{\text{\tiny$ B$}}^2=0$. For the gauge field $A_\mu(x)$,
it is proved as
\begin{align}
 \boldsymbol\delta_{\text{\tiny$ B$}}^2A_\mu(x)
=&
\boldsymbol\delta_{\text{\tiny$ B$}}(\boldsymbol\delta_{\text{\tiny$ B$}}A_\mu(x))=\boldsymbol\delta_{\text{\tiny$ B$}}(\partial_\mu c(x)-ig[A_\mu(x),\,c(x)]_\ast) \nonumber\\
=&ig\partial_\mu(c(x)\ast c(x))
-ig\boldsymbol\delta_{\text{\tiny$ B$}}A_\mu(x)\ast c(x)
-igA_\mu(x)\ast\boldsymbol\delta_{\text{\tiny$ B$}}c(x)\nonumber\\
&+ig\boldsymbol\delta_{\text{\tiny$ B$}}c(x)\ast A_\mu(x)
-igc(x)\ast\boldsymbol\delta_{\text{\tiny$ B$}}A_\mu(x)=0
\end{align}
owing to  the rule \eqref{4.8}. For other fields ,
\begin{align}
\boldsymbol\delta_{\text{\tiny$ B$}}^2{\mit\Psi}(x)=&ig
\boldsymbol\delta_{\text{\tiny$ B$}}(c(x)\ast {\mit\Psi}(x))\nonumber\\
=&ig\boldsymbol\delta_{\text{\tiny$ B$}}c(x)\ast{\mit\Psi}(x)
-igc(x)\ast\boldsymbol\delta_{\text{\tiny$ B$}}{\mit\Psi}(x)=0,\\
\boldsymbol\delta_{\text{\tiny$ B$}}^2c(x)=&
\boldsymbol\delta_{\text{\tiny$ B$}}(igc(x)\ast c(x))\nonumber\\
=&
ig\boldsymbol\delta_{\text{\tiny$ B$}}c(x)\ast c(x)
-igc(x)\ast\boldsymbol\delta_{\text{\tiny$ B$}}c(x)=0,\\
\boldsymbol\delta_{\text{\tiny$ B$}}^2{\bar c(x)}=&
\boldsymbol\delta_{\text{\tiny$ B$}}(iB(x))=0,\\
\boldsymbol\delta_{\text{\tiny$ B$}}^2B(x)=&0.
\end{align}
It should be noted that these BRST transformations  of fields
expressed in the form of representation of nonabelian group
such as in Eqs.\eqref{2.3}, \eqref{4.6}, 
\eqref{4.7} and \eqref{4.8a}
corresponds with those introduced by Soroush \cite{SOR} 
if they are represented in components of fields.
\par
The parameter of BRST transformation is the Grassmann variable $\lambda$
which doesn't depend on $x_\mu$, that means the BRST symmetry is global.
In such a case, the conserved
current $j_\mu(x)$ exists in commutative field theory 
according to the Noether theorem.
However, in the case of noncommutative field theory, it was deduced by Micu and Sheikh-Jabbari
\cite{Micu} that
the divergence of the current is equal to the Moyal $\ast$product of
the some functions.
\begin{equation}
\partial^\mu j_\mu(x)=[f(x),\,g(x)]_\ast,
\end{equation}
where functions $f(x)$ and $g(x)$ are specific to the symmetry.
Here, we would like to indicate the $\ast$commutator to be rewritten in the 
form of the total derivative. It follows that 
\begin{align}
[f(x),\,g(x)]_\ast=& \left(e^{\frac12\theta^{\mu\nu}
\partial_\mu^1\partial_\nu^2}
-e^{-\frac12\theta^{\mu\nu}
\partial_\mu^1\partial_\nu^2}\right)f(x_1)g(x_2)\left.\right|_{x_1=x_2=x}
\nonumber\\
=&\theta^{\mu\nu}\partial_\mu f(x)\partial_\nu g(x)+
\frac{1}{3!}\theta^{\mu\nu}\frac12\theta^{\mu_1\nu_1}\frac12\theta^{\mu_2\nu_2}
\partial_\mu\partial_{\mu_1}\partial_{\mu_2} f(x)\partial_\nu\partial_{\nu_1}\partial_{\nu_2} g(x)+\cdots \nonumber\\
=&\partial^{\mu}\left\{
\theta^{\mu\nu} f(x)\partial_\nu g(x)+
\frac{1}{3!}\theta^{\mu\nu}\frac12\theta^{\mu_1\nu_1}\frac12\theta^{\mu_2\nu_2}
\partial_{\mu_1}\partial_{\mu_2} f(x)\partial_\nu\partial_{\nu_1}\partial_{\nu_2} g(x)+\cdots 
\right\}\nonumber\\
=&\partial^\mu h_\mu(x)
\label{4.21}
\end{align}
because of the antisymmetric $\theta^{\mu\nu}$.
Thus, if the current is redefined to be $J_\mu(x)=j_\mu(x)-h_\mu(x)$, 
$J_\mu(x)$ is the conserved current.
As a result, the Noether theorem may cover the noncommutative field theory.
This is the case also in the nonabelian gauge theory on 
noncommutative spacetime. \par
Let us investigate the BRST current which results from the BRST 
invariance of the Lagrangian of the nonabelian gauge theory 
given by Eq.\eqref{4.1}. From Eq.\eqref{2.22a},
\begin{align}
{\cal L}(A{'}_{\mu}(x),\,{\mit\Psi}{'}(x)&,\,c{'}(x),\,{\bar c}{'}(x),\,
B{'}(x)\,)\nonumber\\
&-{\cal L}(A_\mu(x),\,{\mit\Psi}(x),\,c(x),\,{\bar c}(x),\,
B(x)\,)=ig\lambda\text{Tr}[c(x),\,F_{\mu\nu}\ast F^{\mu\nu}]_\ast,
\label{4.22}
\end{align}
where the prime signs on fields denote BRST transformed fields.
Owing to the equation of motions, the left-hand side of the equation 
changes to
\begin{align}
LHS=&
-\partial_\mu
\left[\rule{0mm}{6mm}\right. 
\sum_{a=0}^{N^2-1}\left\{\rule{0mm}{5mm}\right.
F^{a\,\mu\nu}(x,\theta)\ast 
\delta_{\text{\small$ B$}}A_{\nu}^a(x,\theta)
-i\partial^\mu{\bar c}^a(x,\theta)\ast \delta_{\text{\small$ B$}}c^a(x,\theta)
\nonumber\\
&+i(D^\mu c(x,\theta))^a\ast \delta_{\text{\small$ B$}}{\bar c}^a(x,\theta)
\left.\rule{0mm}{5mm}\right\}
+{\bar{\mit\Psi}}(x,\theta)i\gamma^{\mu}
\ast\delta_{\text{\tiny$ B$}}{\mit\Psi}(x,\theta)
\left.\rule{0mm}{6mm}\right]\nonumber\\
&+\sum_k[\,f_k(x,\theta),\,g_k(x,\theta)\,]_\ast,
\label{4.23}
\end{align}
where the last term  as well as 
equations of motion are explicitly written in Appendix.
Removing the Grassmann parameter $\lambda$ from the left-hand side, 
we may define the BRST current 
\begin{align}
j_\mu=&-
F^{a\,\mu\nu}(x)\ast 
\boldsymbol\delta_{\text{\tiny$ B$}}A_{\nu}^a(x)+{\bar{\mit\Psi}}(x)i\gamma^{\mu}\ast
\boldsymbol\delta_{\text{\tiny$ B$}}{\mit\Psi}(x)\nonumber\\
&-i\partial^\mu{\bar c}^a(x)\ast \boldsymbol\delta_{\text{\tiny$ B$}}c^a(x)
+i(D^\mu c(x))^a\ast \boldsymbol\delta_{\text{\tiny$ B$}}{\bar c}^a(x)
,
\end{align}
where the symbol of the sum over superscript $a$ is abbreviated. 
From Eq.\eqref{4.23}, the equation of continuity of BRST current
\begin{align}
\partial^\mu j_\mu=-\frac{1}{\lambda}\sum_k[\,f_k(x),\,g_k(x)\,]_\ast
\label{4.25a}
\end{align}
follows.
\par
If the last term of Eq.\eqref{4.25a} 
and the right-hand side of Eq.\eqref{4.22}
are rewritten as in Eq.\eqref{4.21}, we may define the current $J_\mu(x)$
which is conserved without any restrictions. However, it is not necessary
to do so because our formulation is Lorentz covariant as in Section 3 and 
the noncommutative parameter $\theta^{\,\mu\nu}$ may be taken to be 
$\theta^{0i}=0$ if the appropriate Lorentz transformation is carried out.
The corresponding BRST charge  
\begin{equation}
Q_B=\int d^3x\, j_0(x)
\label{4.25}
\end{equation}
is conserved since
\begin{equation}
\int d^3x\,[f(x),\,g(x)]_\ast =0
\end{equation}
owing to only space-like noncommutativity $\theta^{0i}=0$.\par
Space-noncommutativity enables us to quantize the fields without difficulties.
The conjugate momentums of fields are defined as 
\begin{align}
&\pi^{a\mu}=\partial {\cal L}/\partial {\dot A}_{\mu}^a=
F^a\rule{0mm}{2.5mm}^{\mu0},\label{4.27}\\
&\pi^{a}_{\text{\tiny$B$}}=\partial {\cal L}/\partial {\dot B}^a=
-A^a_0,\\
&\pi^{k\alpha}_{{\mit\Psi}}=\partial {\cal L}/\partial \,{\dot {\mit\Psi}^{k\alpha}}={\bar{\mit\Psi}}^{k\beta}\,i\gamma^{0\,\alpha}_{\,\beta},\\
&\pi^{a}_{c}=\partial {\cal L}/\partial\, {\dot c}^{\,a}=
-i\,{\dot{\bar c}}^{\,a},\\
&\pi^{a}_{\bar {c}}=\partial {\cal L}/\partial \,{{\dot{\bar{c}}}}^{\;a}=
i\,(D_0c)^{a},\label{4.31}
\end{align}
where $(D_{\mu} c)^a=\partial_\mu{c}^a+gf^{abc}A_\mu\ast c^c-gh^{abc}[A_\mu^b,\,c^c]_\ast$ and dots on fields denote time derivative.
Canonical commutation relations between fields and those conjugate momentums 
follows from the ordinary quantization 
\begin{align}
&[A_j^a({\boldsymbol {x}},t),\,\pi^{bk}({\boldsymbol {y}},t)]
=i\delta^{ab}\delta^k_{j}\delta^3({\boldsymbol {x}}-{\boldsymbol {y}}),
\hskip1cm(j,k=1,2,3),\\
&[A_0^a({\boldsymbol {x}},t),\,\pi^{b}_{\text{\tiny$B$}}({\boldsymbol {y}},t)]
=i\delta^{ab}\delta^3({\boldsymbol {x}}-{\boldsymbol {y}}),\\
&[{\mit\Psi}^{k\alpha}({\boldsymbol {x}},t),\,\pi^{j\beta}
({\boldsymbol {y}},t)]
=i\delta^{kj}\delta^{\alpha\beta}\delta^3
({\boldsymbol {x}}-{\boldsymbol {y}}),\\
&\{c^a({\boldsymbol {x}},t),\,\pi_c^b({\boldsymbol {y}},t)\}=
\{{\bar{c}}^a({\boldsymbol {x}},t),\,\pi_{\bar c}^b({\boldsymbol {y}},t)\}=
=i\delta^{ab}\delta^3({\boldsymbol {x}}-{\boldsymbol {y}}),
\end{align}
which yield together with Eqs.(\ref{4.27})-(\ref{4.31})
\begin{equation}
[i\lambda Q_B,\, {\mit\Phi}_k(x)]=\lambda \boldsymbol\delta_{\text{\tiny$ B$}}
{\mit\Phi}_k(x),\label{4.36}
\end{equation}
where ${\mit\Phi}_k(x)$ represents every related field.
Equation \eqref{4.36} confirms that the charge $Q_B$ is 
a generator of BRST symmetry.
\section{Scale symmetry of ghost fields}
The Lagrangian \eqref{4.1} is invariant under the scale transformation
\begin{equation}
c^a(x)\longrightarrow e^{s}c^a(x),\hskip1cm
{\bar c}^{\,a}(x)\longrightarrow e^{-s}{\bar c}^{\,a}(x),
\label{5.1}
\end{equation}
where $s$ is a real parameter. It should be noted that
ghost fields $c^a$ and ${\bar c}^{\,a}$ are real fields, so that 
they hasn't the phase transformation.
From the invariance of \eqref{4.1} under the transformation \eqref{5.1},
it follows that
\begin{equation}
\partial^\mu i\{{\bar c}\ast (D_\mu c)^a-\partial_\mu{\bar c}^{\,a}\ast c^a\}
=gh^{abc}[i\partial^\mu {\bar c}^{\,a}\ast c^c,\,A_\mu^b]_\ast,
\label{5.2}
\end{equation}
from which the Noether current and its charge are derived
\begin{align}
&j_\mu^{(c)}=i({\bar c}\ast (D_\mu c)^a-\partial_\mu{\bar c}^{\,a}\ast c^a)
,\\
&Q_c=i\int d^3x\, ({\bar c}^{\,a}\ast (D_0 c)^a
-\partial_0{\bar c}^{\,a}\ast c^a).
\label{5.4}
\end{align}
$Q_c$ is called the FP ghost charge and  conserved owing to 
the space-like noncommutativity $(\theta^{0i}=0)$ and Eq.\eqref{5.2}.
$Q_c$ is also a generator of scale transformation Eq.\eqref{5.1} and satisfies
that
\begin{equation}
[\,iQ_c,\,c^a(x)\,]=c^a(x),\hskip1cm 
[\,iQ_c,\,{\bar c}^{\,a}(x)\,]=-{\bar c}^{\,a}(x),
\end{equation}
which deduces that the number operator of ghost field $N_{F\!P}$ is 
defined as
\begin{equation}
N_{F\!P}=iQ_c.
\end{equation}
\par
According to the definitions Eqs.\eqref{4.25} and \eqref{5.4}, 
the BRST charge and FP ghost charge satisfy
the following simple algebra.
\begin{align}
&\{Q_{B},\,Q_{B}\}=2Q_{B}^2=0,\\
&[iQ_{c},\,Q_{B}]=Q_{B},\\
&[Q_{c},\,Q_{c}]=0,
\end{align}
which may 
play an important role in the classification of asymptotic fields and 
the proof of S matrix unitarity.
\section{conclusions}
The deviation from the standard model in particle physics has not yet observed,
and so any model beyond standard model must reduce to it in 
some approximation. Noncommutative gauge theory must also reproduce standard
model in the limit of noncommutative parameter $\theta^{\mu\nu}\to0$. 
In order to meet this condition,
we have formulated following to Jur$\check{\text{c}}$o {\it et. al.} Jurco
 the nonabelian gauge theory on the noncommutative 
spacetime which reduces to the ordinary gauge theory in the limit 
$\theta^{\mu\nu}\to0$. This is because our noncommutative gauge theory 
depends on the enveloping group SU(N)$\ast$ that is made of the enveloping
Lie algebra with the condition Eq.\eqref{2.13b}. However, 
It should be investigated
if the noncommutative gauge theory thus constructed may clear the 
allegations pointed out by Armoni\cite{Armoni}, which is our future work. 
\par
Lorentz covariance is the fundamental principle of every relativistic field theory which insures consistent physical descriptions. Even if the space-time is noncommutative, field theories on it should keep Lorentz covariance.
In this paper, we defined the nonabelian gauge theory on noncommutative spacetime where its Lorentz covariance is maintained since 
one specific 
eigen state of the noncommutative operator ${\hat\theta}^{\mu\nu}$ 
is  picked up in the 6 dimensional $\theta$ space.
We can choose the spacelike noncommutativity ($\theta^{0i}=0$) after 
the appropriate Lorentz transformation and so any serious outcomes such as 
quantization and unitarity violation don't appear.
\par
We also investigated the BRST symmetry of noncommutative nonabelian 
gauge theory. Our formulation is more transparent that that by 
Soroush \cite{SOR} because 
we consider the BRST transformations of the fields 
as the representations of gauge group, not component of fields 
as in \cite{SOR}. The scale transformation of ghost fields is also
investigated. 
BRST symmetry as well as the scale transformation
  are very important to quantize 
the ordinary 
gauge theory and achieve the covariant canonical formulation of it. 
The Ward-Takahashi identify derived from BRST symmetry plays an important
role in the proof of renormalizability of the gauge theory.
It could be certainly
important in the case of the nonabelian gauge theory 
on noncommutative spacetime. The detailed study about these interesting 
subjects is our future work.

\appendix
\section{equation of motion }
\begin{align}
&\partial_\mu F^a\rule{0mm}{2.8mm}^{\mu\nu}+gf^{abc}A_\mu^b\ast
F^c\rule{0mm}{2.8mm}^{\mu\nu}
+g\,{\bar{\mit\Psi}}\gamma^\nu T^a\ast{\mit\Psi}
-igf^{abc}\partial^{\,\nu}{\bar c}^{\,b}\ast c^{\,c}
-\partial^{\,\nu} B^a\nonumber\\
&\hskip2cm=
gh^{abc}[A_\mu^b,\,F^c\rule{0mm}{2.8mm}^{\mu\nu}]_\ast
-ig(f^{abc}-h^{abc})\{\partial^\nu{\bar c}^{\,b},\,c^{\,c}\}_\ast
+g\{{\bar{\mit\Psi}}^\mu\rule{0mm}{2.8mm}^i,\,{\mit\Psi}^{\beta}\rule{0mm}{2.8mm}^{j}\}_\ast \gamma^\nu_{\mu\beta}T^a_{ij},\\
&(i\gamma^\mu {\cal D}_\mu-m){\mit\Psi}=0,\\
&\partial^\mu D_\mu c^a=0,\\
& (D^\mu \partial_\mu{\bar c})^{\,a}+gh^{abc}[\partial^\mu {\bar c}^{\,b},\,
A_\mu^{\,c}]_\ast=0,\\
&\partial^\mu A_\mu^a+B^a=0,
\end{align}
where space term coordinate $x_\mu$ is abbreviated and 
$h^{abc}=\frac12(f^{abc}+id^{abc})$.
\section{$\ast$Commutator in Noether theorem}
\begin{align}
\sum_k[\,f_k(x),\,g_k(x)\,]_\ast
=&-\frac12[\,\delta_{\text{\tiny$ B$}}(\partial_\mu A_\nu^a),
\,{F^a}\rule{0mm}{2.5mm}^{\mu\nu}\,]_\ast\nonumber\\
&-\frac18g(f^{abc}-id^{abc})\left\{\rule{0mm}{3.22mm}\right.
[A_\mu^b\ast \delta_{\text{\tiny$ B$}}A_\nu^c,\,
{F^a}\rule{0mm}{2.5mm}^{\mu\nu}\,]_\ast\nonumber\\
&+[\,\delta_{\text{\tiny$ B$}}A_\mu^b,\,A_\nu^c\ast 
{F^a}\rule{0mm}{2.5mm}^{\mu\nu}\,]_\ast
+[\,{F^a}\rule{0mm}{2.5mm}^{\mu\nu}\ast 
\delta_{\text{\tiny$ B$}}A_\mu^b,\,A_\nu^c\,]_\ast
\left.\rule{0mm}{3.22mm}\right\}\nonumber\\
&-\frac14gh^{abc}\left\{\rule{0mm}{3.22mm}\right.
[A_\nu^c\ast \delta_{\text{\tiny$ B$}}A_\mu^b,\,
{F^a}\rule{0mm}{2.5mm}^{\mu\nu}\,]_\ast\nonumber\\
&+[\,\delta_{\text{\tiny$ B$}}A_\nu^c,\,A_\mu^b\ast 
{F^a}\rule{0mm}{2.5mm}^{\mu\nu}\,]_\ast
+[\,{F^a}\rule{0mm}{2.5mm}^{\mu\nu}\ast 
\delta_{\text{\tiny$ B$}}A_\nu^c,\,A_\mu^b\,]_\ast
\left.\rule{0mm}{3.22mm}\right\}\nonumber\\
&+g\{{\bar{\mit\Psi}}^{i\alpha}\gamma^{\mu}_{\alpha\beta}
\delta_{\text{\tiny$ B$}}A_\mu^a,\,T^a_{ij}\ast {\mit\Psi}^{j\beta}
\}_\ast\nonumber\\
&-ig(f^{abc}-h^{abc})\{\partial^{\mu}{\bar c}^a\ast
\delta_{\text{\tiny$ B$}}A_\mu^b,\,c^c\}_\ast\nonumber\\
&-i[\partial^\mu\delta_{\text{\tiny$ B$}}{\bar c}^a,\,(D_\mu c)^a]_\ast
-igh^{abc}[\partial^\mu{\bar c}^a\ast 
\delta_{\text{\tiny$ B$}}{\bar c}^c,\,A_\mu^b]_\ast
\end{align}


\end{document}